\documentclass[pra,twocolumn,superscriptaddress,showpacs]{revtex4}
\usepackage{amsmath,amssymb,amsfonts}
\usepackage{graphicx,color}
\begin{document}

\newcommand\calA{\mathcal{A}}
\newcommand\calH{\mathcal{H}}
\newcommand\Tr{\operatorname{Tr}}
\newcommand\opt{\mathrm{opt}}
\newcommand\sign{\operatorname{sgn}}

\title{Local and Nonlocal Contents in N-qubit generalized GHZ states}
\author{Chang-liang Ren}
\affiliation{Department of Physics, Korea University, Seoul
136-713, Korea}
\author{Mahn-Soo Choi}
\email{choims@korea.ac.kr}
\affiliation{Department of Physics, Korea University, Seoul 136-713, Korea}

\begin{abstract}
We investigate local contents in $N$-qubit generalized Greenberger, Horne, and
Zeilinger (GHZ) states.
We suggest a decomposition for correlations in the GHZ states into a nonlocal
and fully local part, and find a lower and upper bound on the local content.
Our lower bound reproduces the previous result for $N=2$ [Scarani,
Phys. Rev. A. 77, 042112 (2008)] and decreases rapidly with $N$.
\end{abstract}

 \maketitle

\section{Introduction}
\label{paper::sec:1}

Bell's theorem has revealed that local variable theories
cannot reproduce all statistical predictions of quantum theory,
and
highlights the statistical incompatibility between classical local variable
theories and quantum theory \cite{Bell,CHSH}.  When the Bell-type inequalities
are violated, non-locality appears.
However, even if the observations on a given system of particle pairs exhibit
non-locality, it does not necessarily imply that all individual pairs in the
system behave non-locally.
It may be possible that some fraction of the pairs behave non-locally, while
the other pairs behave locally.

This issue has been investigated carefully first by Elitzur, Popescu, and
Rohrlich (to be referred to as EPR2) \cite{EPR2} in terms of the local contents
in a given non-local correlation.
Since then, several authors generalized and further discussed the idea
\cite{Brunner,Scarani,Almeida,Barrett,Zhang,Branciard}.  For example, EPR2
approach has been related to another noticeable question, the simulation of
quantum correlations with other resource, which has been proved useful in the
task of simulating entanglement \cite{Brunner}.
Barrett \textit{et al.} \cite{Barrett} gave an upper bound of the weight of
local component in $d\times d$ system .  Scarani \cite{Scarani} presented an
improved lower and upper bound of the local content in the family of pure
two-qubit states and the first example of a lower bound of the local content
in pure two-qutrit system. Later, Zhang \textit{et al.} \cite{Zhang} extended
this lower bound to the mixed two-qubit states. In \cite{Branciard}, a new
EPR2 decomposition has been given, which can make their local content reach
the upper bound in a wide range of two-qubit pure states.

In this paper, we investigate local contents in $N$-qubit generalized
Greenberger, Horne, and Zeilinger (GHZ) states.  In
Section~\ref{paper::sec:2}, we first examine general
requirements for optimization of the weight of the local part in a convex
decomposition of a given quantum distribution into local and non-local part.
Guided by the requirements, we suggest local probability distributions for
$N$-qubit GHZ states, which gives the lower bound of the local contents of
such states, in Section~\ref{paper::sec:3}.  Our result for $N=2$ reproduces
the lower bound in \cite{Scarani}. It is also noted that the lower bound on
the local content decreases rapidly with $N$.  In Section~\ref{paper::sec:6},
we briefly discuss the upper bound based on Bell-type inequalities.

\section{EPR2 Approach}
\label{paper::sec:2}

Before we go further, here we first review the notion of \emph{local content}
suggested first by Elitzur, Popescu, and Rohrlich \cite{EPR2}.  We follow the
conventions in Ref.~\cite{Scarani}.

Consider a system of $N$ parties, labeled by $1, 2, . . . ,N$.
On each party $j$, one measures any observable $A_j$ in a given set
$\calA_j$.  The measurement output of $A_j$ is denoted by $r_j$.
The joint probability distribution for measurements on the system
is denoted by
\begin{math}
P(r_{1},r_{2},...,r_{N}|A_{1}, A_{2},...,A_{N}).
\end{math}
Under the circumstance that the parties are non-communicating but
share classical information,
the joint probability distributions takes the following form:
\begin{multline}
\label{Local}
P(r_{1},r_{2},...,r_{N}| A_{1}, A_{2},...,A_{N})= \\
\int d\mu(\lambda)P(r_{1}|A_{1},\lambda)P(r_{2}\mid
A_{2},\lambda)...P(r_{N}|A_{N},\lambda).
\end{multline}
where $\lambda\in \Lambda$ denotes the collective local hidden variables that
represent the shared classical information and $\Lambda$ is the space of all
hidden variables. The form of the distribution in \eqref{Local} leads to a set
of constraints on the joint distributions (Bell-type inequalities) for any
fixed number of measurements on each party. If there exist joint probability
distributions that violates the inequalities, they would not be written as
\eqref{Local}, and are thus non-local.

The quantum correlations are obtained by general measurements on quantum
states, and the joint probability distributions is given by
\begin{multline}
P_{Q}(r_{1},r_{2},...r_{N}|A_{1},
A_{2},...,A_{N};\rho)= \\ \Tr\left(\Pi_{r_{1}}^{A_{1}}\otimes
\Pi_{r_{2}}^{A_{2}}\otimes\cdots\otimes
\Pi_{r_{N}}^{A_{N}}\rho\right).
\end{multline}
Here $\rho$ is the density matrix for a quantum state of the system of $N$
parties.  $\Pi_{r_{j}}^{A_{j}}$ is the projector on the subspace associated to
the measurement result $r_{j}$ of the observable $A_{j}$ performing on party
$j$. There exist quantum probability distributions that are not local, as
proved by Bell \cite{Bell}.

EPR2 approach is a quantitative notion of non-locality \cite{EPR2}. The main
idea is to consider the possible decomposition of $P_{Q}$ into a local part
$P_{L}$ and a nonlocal part $P_{NL}$:
\begin{equation}
\label{paper::eq:1}
P_{Q}=w(\rho)P_{L}+[1-w(\rho)]P_{NL}
\end{equation}
where the weight $w\in [0,1]$ of the local component is required to be
independent of the measurements and the outcomes. Obviously, the convex
combination \eqref{paper::eq:1} is not unique.  The point is to find the local
part $P_L$ that maximizes the weight $w$.  The resulting optimal value
$w_\opt$ of $w$ is defined as the \emph{local content} in the joint
probability distribution distribution $P_Q$.  The local content $w_\opt(\rho)$
should be $1$ if $\rho$ is a product state, and $0$ if $\rho$ is a maximally
entangled state \cite{EPR2,Barrett}.

The full optimization of the local part $P_L$ is highly non-trivial.  Several
authors have investigated the local contents in two-qubit and two-qutrit
states, and proposed upper and lower bounds on the local contents \cite{EPR2,
  Scarani, Barrett, Zhang, Branciard}.  Here we intend to give lower and upper
bounds of the local content $w_\opt$.  in the $N$-qubit generalized GHZ states
of the form
\begin{equation}
\label{paper::eq:3}
\left\vert \Psi_{n}(\alpha) \right\rangle=\cos\alpha\left\vert 0...0
\right\rangle+\sin\alpha\left\vert 1...1 \right\rangle
\end{equation}
where $\alpha\in[0,\pi/4]$.

\section{Lower Bound on the Local Content}
\label{paper::sec:3}

Here we provide the lower bound $w_\opt^<$ of the local content $w_\opt$ by
finding reasonable local probability distribution function $P_L$ guided by the
following requirements: (i) As the the non-local part $P_{NL}$ is a
probability distribution and non-negative,
\begin{multline}
P_{Q}(r_{1},r_{2},...r_{N}|A_{1}, A_{2},...,A_{N};\rho) \geq \\
wP_{L}(r_{1},r_{2},...,r_{N}|A_{1},A_{2},...,A_{N})
\end{multline}
for all possible local measurements $A_j$
and outcomes $r_j$.
In particular, $P_L$ should be zero whenever $P_Q$ is zero.
(ii) As $P_{L}(r_{1},r_{2},...,r_{N}|A_{1}, A_{2},...,A_{N})$
is a real probability distribution,
\begin{equation}
\sum_{r_{1},r_{2},...,r_{N}}P_{L}(r_{1},r_{2},...,r_{N}|A_{1},
A_{2},...,A_{N})=1
\end{equation}

For an arbitrary $N$-qubit state $\rho_N$, the
joint probability distribution is given by
\begin{multline}
P_{Q}(r_{1},\cdots,r_{N}|A_{1},\cdots,A_{N};\rho_{N}) = \\{}
\Tr\left(\Pi_{r_{1}}^{A_{1}}\otimes\cdots\otimes
\Pi_{r_{N}}^{A_{N}}\rho_N\right)
\end{multline}
with the the projectors defined by
\begin{eqnarray}
\Pi_{r_{j}}^{A_{j}}=\frac{1}{2}(I+r_{i}\,\vec{n}_{i}\cdot\vec\sigma)
\end{eqnarray}
where $r_{1},r_{2} =\pm 1$, $\vec\sigma$ denotes the three Pauli matrices, and
$\vec{n}_{i}=(\sin\theta_{i}\cos\varphi_{i}, \sin\theta_{i}\sin\varphi_{i},
\cos\theta_{i})$.
Without loss of generality,
by readjusting the quantization axis if necessary,\cite{endnote:1}
we assume
that
\begin{equation}
\varphi_1+\cdots+\varphi_N=\pi \,.
\end{equation}
Then the quantum joint probability distribution corresponding to the $N$-qubit
GHZ state in Eq.~(\ref{paper::eq:3}) can be written as
\begin{multline}
\label{paper::eq:2}
P_{Q}=\frac{\cos^2\alpha}{2^{N}}\prod_{j=1}^N(1+r_{j}\cos\theta_{j})  +
\frac{\sin^2\alpha}{2^N}\prod_{j=1}^N(1-r_j\cos\theta_j) \\{}
- \frac{\sin(2\alpha)}{2^N}\prod_{j=1}^N r_j\sin\theta_j
\end{multline}

\subsection{$N=2$ Case}

In their original paper \cite{EPR2}, EPR2 proposed an explicit local
probability distribution $P_{L}$, which leads to a decomposition of the
form in ~(\ref{paper::eq:1}) with
\begin{math}
w(\alpha)=[1-\sin(2\alpha)]/4.
\end{math}
This is the first known lower bound on $w_\opt(\alpha)$. They proved that the
bound is tight for the maximally entangled state and under a reasonable
continuity assumption; the singlet state of two qubits is fully
nonlocal. However, for the product state which is fully local, $w$ equals to
$1/4$ instead of $1$. So this decomposition is not optimal.  Later, Scarani
suggested a modified explicit local probability distribution $P_{L}$, which
can lead to an EPR2 decomposition with $w_\opt^<(\alpha)=1-\sin(2\alpha)$
\cite{Scarani}.

Here we exploit a method to find local distribution function $P_L$, which can
be easily extended to the cases with $N>2$.  In order to optimize $w$ in the
decomposition (\ref{paper::eq:1}) as much as possible, we take a note of the
requirement as discussed in Section~\ref{paper::sec:2} that $P_{L}=0$ whenever
$P_{Q}=0$.  and that $P_{L}$ should approach $P_{Q}$ as much as possible.  In
the special case of
\begin{math}
r_1=r_2=1
\end{math}
and
\begin{math}
\theta_1=\theta_2=\theta,
\end{math}
the quantum probability distribution in \eqref{paper::eq:2} reduces to
\begin{equation}
\label{Q21}
P_{Q}=\frac{1}{4}\left\{
  2\cos\theta(1+\cos(2\alpha)]
  -\sin^2\theta[1+\sin(2\alpha)]\right\}.
\end{equation}
We note that $P_Q=0$ only when $\theta=\theta_0$, where
\begin{eqnarray}
\label{Q22}
\cos\theta_0\equiv -\frac{1-\tan\alpha}{1+\tan\alpha}
\end{eqnarray}
That means that we must have $P_{L}=0$ at $\theta=\theta_0$. Besides $P_{L}$
should approach $P_{Q}$ as close as possible.  We thus suggest a local
probability distribution $P_{L}$ of the form
\begin{multline}
\label{paper::eq:4}
P_{L}=\frac{1}{4}\left[1+\sign(\cos\theta_{1})
  \min\left(1,\left|\frac{\cos\theta_1}{\cos\theta_0}\right|\right)
\right]\\{}\times
\left[1+\sign(\cos\theta_{2})
  \min\left(1,\left|\frac{\cos\theta_2}{\cos\theta_0}\right|\right)\right].
\end{multline}
It is easy to see that this form can assure that $P_{L}=0$ whenever $P_{Q}=0$:
Obviously, in special situation that $ \theta_{1}=\theta_{2}=\theta$, if
$P_{Q}=0$, $P_{L}$ is zero.
In a general situation that $\theta_{1}\neq
\theta_{2}$ and $P_{Q}=0$,
it follows from the form in \eqref{paper::eq:2} that $P_Q=0$ at
$(\theta_1,\theta_2)$ such that either $\cos\theta_{1}>
\cos\theta_{0}>\cos\theta_{2}$ or $\cos\theta_{2}>
\cos\theta_{0}>\cos\theta_{1}$. When $\cos\theta_{1}>
\cos\theta_{0}>\cos\theta_{2}$, the second factor in Eq.~(\ref{paper::eq:4})
vanishes, and vice versa.

Previously, we discussed the situation that $r_{1}=r_{2}=1$, but a
valid local probability distribution should contain all local
measurements and outcomes. So we give the complete local probability
distribution $P_{L}$ as
\begin{equation}
\label{PL22}
P_{L}=\frac{1}{4}\prod_{j=1}^2\left[1+r_j\sign(\cos\theta_{j})
  \min\left(1,\left|\frac{\cos\theta_j}{\cos\theta_0}\right|\right)
\right].
\end{equation}
Note that this form of the local distribution function is identical to the one
in Ref.\cite{Scarani}
\begin{math}
(\frac{1+\tan\alpha}{1-\tan\alpha}=\frac{\cos2\alpha}{1-\sin2\alpha}).
\end{math}
Once the local component $P_L$ is fixed, the weight $w(\alpha)$ is optimized
to give the lower bound $w_\opt^<$ on the local content by minimizing the
function $f(\theta)$, defined by
\begin{equation}
\label{paper::eq:5}
f(\theta)\equiv\frac{P_{Q}(\theta)}{P_{L}(\theta)}
\end{equation}
where $P_Q(\theta)$ and $P_L(\theta)$ is the quantum and local joint
probability functions, respectively, in the special case
$\theta_1=\theta_2=\theta$.
For the present case ($N=2$),
\begin{equation}
f(\theta)
=\frac{\left[1+2\cos(2\alpha)\cos\theta
    +\cos^2\theta-\sin(2\alpha)\sin^2\theta\right]}{
  \left[1+\sign(\cos\theta)
  \min\left(1,\left|\frac{\cos\theta}{\cos\theta_{0}}\right|\right)\right]^2}.
\end{equation}
The local distribution function proposed in (\ref{PL22}) allows the resulting
lower bound $w_\opt^<$ to reach $1-\sin(2\alpha)$ obtained previously by
Scarani\cite{Scarani}. The profile of the lower bound $w_\opt^<(\alpha)$ of
local content vs $\alpha$ is shown in Fig.~\ref{paper::fig:1}
($N=2$). Clearly, $w_\opt^<(\alpha)$ decreases with $\alpha$ eventually
vanishing at $\alpha=\pi/4$, as it should since the degree of entanglement in
the GHZ state in Eq.~(\ref{paper::eq:3}) increases with $\alpha$ reaching the
maximal entanglement at $\alpha=\pi/4$.

\subsection{$N=3$ Case}

\begin{figure}
\begin{center}
\includegraphics[width=6cm]{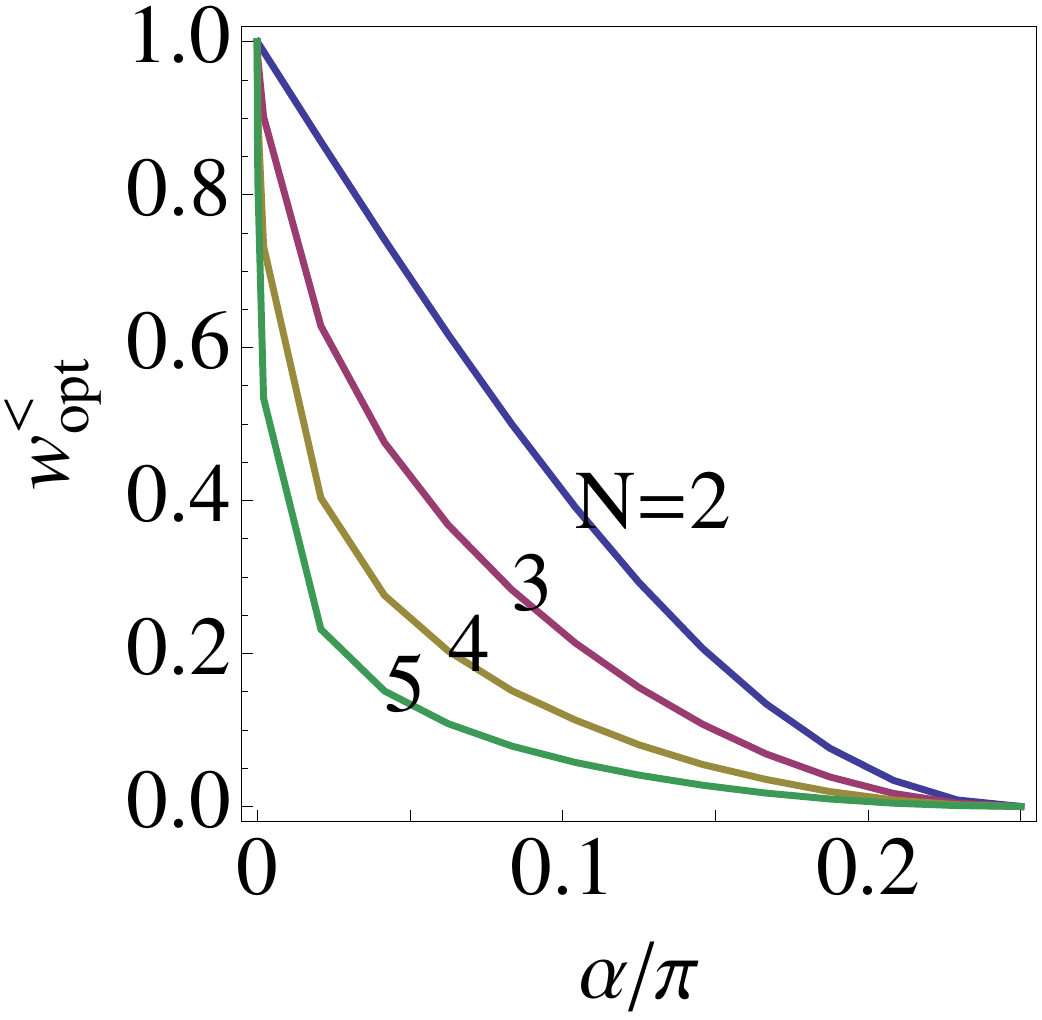}
\end{center}
\caption{The lower bound $w_\opt^<$ on the local content $w_\opt$ in pure
  $N$-qubit generalized GHZ states as a function of $\alpha$ for $N=2,3,4,5$
  (top to bottom).}
\label{paper::fig:1}
\end{figure}

As before, we first consider the special situation $r_{1}=r_{2}=r_{3}=1$ and
$\theta_{1}=\theta_{2}=\theta_{3}=\theta$, where the quantum probability
distribution~(\ref{paper::eq:2}) is reduced to to
\begin{equation}
\label{Q31}
P_{Q}=\frac{1}{16}\left[
  \cos(\alpha-3\theta/2)+3\cos(\alpha+\theta/2)
\right]^2.
\end{equation}
$P_Q=0$ only when $\theta=\theta_0$, where
\begin{equation}
\label{Q32}
\cos\theta_0\equiv
-\frac{1-\tan^{\frac{2}{3}}\alpha}{1+\tan^{\frac{2}{3}}\alpha}
\end{equation}
Following the same lines as in the case of $N=2$,
we suggest for the local probability distribution
\begin{equation}
\label{PL33}
P_{L}=\frac{1}{8}\prod_{j=1}^3\left[1+r_j\sign(\cos\theta_{j})
  \min\left(1,\left|\frac{\cos\theta_j}{\cos\theta_0}\right|\right)
\right].
\end{equation}
Given the form of local distribution function~(\ref{PL33}), the lower bound on
the local content is again determined by minimizing the function $f(\theta)$
in Eq.~(\ref{paper::eq:5}). 
Note that unlike the previous case, the lower bound of local content can not
reach $1-\sin(2\alpha)$ except for the maximally entangled state and the
product state. For example, when $\alpha=\frac{\pi}{12}$, $w_\opt^<=0.28$.
The profile of $w_\opt^<(\alpha)$ vs $\alpha$ is shown in
Fig.~\ref{paper::fig:1} ($N=3$). As in the previous case with $N=2$, the lower
bound on local content decreases with $\alpha$ and vanishes at
$\alpha=\pi/4$. It is interesting to note that the lower bound on local
content decreases faster in this case than for $N=2$.  As we will see below
for $N$-qubit states, this trend is general and our lower bound on local
contents in the GHZ state~(\ref{paper::eq:3}) decreases rapid with $N$.

\subsection{General $N$-Qubit Case}

Following the similar lines as above, we define $\theta_0$ by
\begin{equation}
\label{QN2}
\cos\theta_0\equiv
-\frac{1-\tan^{\frac{2}{N}}\alpha}{1+\tan^{\frac{2}{N}}\alpha} \,,
\end{equation}
at which $P_Q=0$ only when $\theta_1=\cdots=\theta_N=\theta_0$.  We then
suggest the following form of the local probability distribution $P_{L}$:
\begin{equation}
P_{L}=\frac{1}{2^N}\prod_{j=1}^N\left[1+r_j\sign(\cos\theta_{j})
  \min\left(1,\left|\frac{\cos\theta_j}{\cos\theta_0}\right|\right)
\right].
\end{equation}
It is interesting to note that the resulting lower bound $w_\opt^<$ on the
local content decreases rapidly with $N$.  Its profile vs $\alpha$ for
$N=2,3,4,5 $ is shown in Fig.~\ref{paper::fig:1}.

\section{Upper Bound on the Local Content}
\label{paper::sec:6}

So far we have focused on the lower bound $w_\opt^<(\rho)$ on the local
content $w_\opt(\rho)$.  Let us now briefly discuss the upper bound
$w_\opt^>(\rho)$.  As pointed out in Ref.~\cite{Barrett}, any Bell inequality
can lead to an upper bound on $w_\opt(\rho)$.  Following
Refs.~\cite{Barrett,Scarani}, suppose that a Bell inequality $P\leq P_L^*$
with a constant $P_L^*$ holds for all local probability distributions. Let
$P_{NS}^*$ be the maximum value of $P$ under the condition of non-signaling
condition.  Then by Eq.~(\ref{paper::eq:1}) and the Bell inequality
\begin{math}
P_Q^*(\rho) \leq w_\opt(\rho)P_L^* + [1-w_\opt(\rho)]P_{NS}^*
\end{math}
where $P_Q^*$ is the quantum value of $P$ for the best choice of measurements.
That is, the upper bound of the local content is given by
\begin{equation}
w_\opt^> = \frac{P_{NS}^*-P_Q^*}{P_{NS}^*-P_L^*}
\end{equation}

An upper bound $w_\opt^>$ for $N=2$ was obtained based on this method in
Ref.~\cite{Scarani}.  Here we thus focus on the case of $N\geq 3$, where a
Bell-type inequality was derived and shown to be violated maximally by GHZ
states in Ref.~\cite{Chenkai}. One can show that $I_{L}^*=1$,
$P_{NS}^*=2^{N-2}$,
\begin{math}
P_Q^*=\sqrt{2^{N-2}\sin^{2}(2\alpha)+\cos^{2}(2\alpha)}.
\end{math}
It immediately follows that
\begin{equation}
w_\opt^> = 
\frac{2^{N-2}-\sqrt{2^{N-2}\sin^{2}(2\alpha)+\cos^{2}(2\alpha)}}{2^{N-2}-1}.
\end{equation}
Obviously, for product states ($\alpha=0$), the upper bound reaches 1, which
is optimal.  However, for the maximally entangled state ($\alpha=\pi/4$), the
upper bound is not optimal, and approaches $1$ as
$N\to\infty$
(to be compared with the result in the bipartite case, $N=2$, in
Ref.\cite{Barrett}).

One can also consider the upper bound based on the Mermin-Ardehali-Belinskii-
Klyshko (MABK) inequality \cite{MABK} for the GHZ states (\ref{paper::eq:3}).
In this case, we find that the upper bound is $0$ for the maximally entangled
state. However, for the generalized GHZ states such that
$\sin(2\alpha)\leq1/\sqrt{2^{N-1}}$, the upper bounds based on the MABK
inequality are $1$ again.  This is because such states do not violate MABK
inequalities \cite{Scarani2}.

\section{Conclusion}
\label{paper::sec:4}

In this paper, we have provided a decomposition for correlations in N-qubit
generalized GHZ states into a nonlocal and fully local part. A general form of
non-trivial local probability distribution $P_{L}$ of $N$ qubits has been
proposed based on the properties of the convex decomposition of the quantum
joint probability distribution into local and non-local parts, and thereby a
lower bound on the local content in the GHZ states has been suggested.  The
improved local probability distribution in \cite{Scarani} for pure two-qubit
states turns out to be a special case of our results.
Moreover, for a fixed value of $\alpha$, our lower bound on the local content
decreases rapidly with $N$.
We have also investigated the upper bound on the local content based on
Bell-type inequalities.

\begin{acknowledgments}
This work was supported by the NRF Grant (2009-0080453), the BK21, the APCTP,
and the KIAS.  C.-L.R.\ is grateful to Prof.~V.~Scarani for helpful
discussions.
\end{acknowledgments}


\begin{thebibliography}{99}
\bibitem{endnote:1} Given a choice of the quantization axes, let $\varphi_j$
($j=1,\cdots,N$) be the azimuthal angle of the $j$th qubit, in general, with
$\sum_j\varphi_j\neq\pi$.  Then rotate the axes counterclockwise around $z$
axis by angle $\varphi_0=(\sum_j\varphi_j-\pi)/N$ so that
$\varphi_j\to\varphi_j'=\varphi_j-\varphi_0$ and hence $\sum_j\varphi_j'=\pi$.
\end{thebibliography}
\end{document}